\documentclass[11pt]{article}

\begin{document}

\title{Variation of $G$, $\Lambda_{(4)}$ and Vacuum Energy From Brane-World Models}
\author{J. Ponce de Leon\thanks{E-mail: jponce@upracd.upr.clu.edu or jpdel1@hotmail.com}\\ Laboratory of Theoretical Physics, Department of Physics\\ 
University of Puerto Rico, P.O. Box 23343, San Juan, \\ PR 00931, USA} 
\date{November 2002}

\maketitle
\begin{abstract}
In brane-world theory in five dimensions, the bulk metric is  usually written in gaussian coordinates, where $g_{4\mu } = 0$ and $g_{44} = - 1$. 
However, the  choice $g_{44} = - 1$ is an external condition, not a requirement of the field equations. 
In this paper we study the consequences of having $g_{44} = \epsilon \Phi^2$, where $\epsilon = \pm 1$ and $\Phi$ is a scalar function varying with time, $\dot{\Phi} \neq 0$. This varying field  entails the  possibility  of variable fundamental physical ``constants". These variations are different from those predicted in scalar-tensor and multidimensional theories.
We solve the five-dimensional equations for a {\em fixed} brane and use the brane-world paradigm to determine the fundamental parameters in the theory, which are the vacuum energy $\sigma$, the gravitational coupling $G$ and  the cosmological term $\Lambda_{(4)}$. We present specific models where these physical quantities are variable functions of time. Different scenarios are possible but we discuss with some detail a model for which $\dot{G}/G \sim H$ and $\Lambda_{(4)} \sim H^2$, which seems to be favored by observations. Our results are not in contradiction to previous ones in the literature. In particular, to those where the brane is described as a domain wall moving in a static $Sch-AdS$ bulk. Indeed these latter models in RS scenarios describe the same spacetime as other solutions (with fixed brane) in gaussian coordinates with $\dot{\Phi} = 0 $.  We conclude that the introduction of a time-varying $\Phi$ in brane-world theory yields a number of models that show variation in the fundamental physical ``constants" and exhibit  reasonable physical properties.   
 
 \end{abstract}



\newpage

\section{INTRODUCTION}

Recently, there has been an increased interest in models where our four-dimensional universe is embedded in a higher-dimensional bulk spacetime having large extra dimensions. The scenario in these models is that matter fields are confined to our four-dimensional universe, a $3$-brane, in a $1 + 3 + d$ dimensional spacetime, while gravity propagates in the extra $d$ dimensions as well. Such scenario was introduced as providing possible solutions to the hierarchy and the cosmological constant problems \cite{RS1}-\cite{RS2}.  

Much work has been done for $d = 1$, where our $3$-brane is a domain wall in a five-dimensional anti-de Sitter spacetime \cite{Maartens1}-\cite{Maartens4}. Since the extra dimension is not compactified, the bulk metric is allowed to be a function of the extra coordinate.   
The general bulk metric ansatz in brane-world models is 
\begin{equation}
\label{usual ansatz}
d{\cal S}^2 = g_{\mu\nu}(x^{\rho}, y)dx^{\mu}dx^{\nu} - dy^2,
\end{equation}
where $x^{\rho} = (x^0, x^1, x^2, x^3)$ are the coordinates in $4D$ and $y$ is the coordinate along the extra dimension, which is assumed to be spacelike. We will use signature $(+ - - -)$ for the four-dimensional space-time, and follow Landau and Lifschitz \cite{Landau and Lifshitz} for the definitions of tensor quantities. 

The effective equations for gravity on a $3$-brane were obtained by Shiromizu {\it et al} \cite{Shiromizu} by using Israel's boundary conditions and imposing ${\bf Z}_2$ symmetry for the bulk spacetime, about our brane {\em fixed} at some $y = y_{0}$. These equations predict five-dimensional corrections to the usual general relativity in $4D$ and provide explicit expressions that link the tension of the brane to the observed gravitational coupling $G$ and cosmological term $\Lambda_{(4)}$. 

Cosmological solutions in the brane-world have been obtained \cite{Binetruy}-\cite{Vollick}  by explicitly solving  the $5D$ field equations in  gaussian normal coordinates (\ref{usual ansatz}). In this approach the junction conditions with the appropriate ${\bf {Z}}_2$ symmetry are solved keeping the brane fixed at some $y = const$. An important feature, relevant to the present work, is that in all these brane-world models the tension of the brane, as well as $G$ and $\Lambda_{(4)}$, can be taken as constants. 

We note that in $5D$, the freedom of coordinates allows us to impose  $5$ conditions on a general five-dimensional metric $g_{AB}$.  The ansatz (\ref{usual ansatz}) makes use of all of them to set $g_{4\alpha} = 0$ and $g_{44} = - 1$. In this work we ask what would happen if we relax some of these conditions. Here we will  consider a variable $g_{44}$. Specifically, we consider the metric 
\begin{equation}
\label{metric with g44 not 1}
d{\cal S}^2 = g_{\mu\nu}(x^{\rho}, y)dx^{\mu}dx^{\nu} + \epsilon \Phi^2(x^{\rho}, y) dy^2,
\end{equation} 
where $\epsilon = -1$ or $\epsilon = + 1$ depending on whether the extra dimension is spacelike or timelike, respectively. 

In this paper, we study the effects of the extra scalar field $\Phi$ on the dynamics in $4D$, within the context of brane-world models. We assume that the brane is fixed at some $y = const$ and find that a variable scalar field $\Phi$ allows the construction of a number of models where, without internal contradictions,  the vacuum energy $\sigma$, the gravitational coupling $G$, and the cosmological term $\Lambda_{(4)}$ have to be variable.  

These results are not in contradiction to well known results in the literature where $\Phi \neq 1$ and still there is no variation in these physical quantities. What we have in mind here is the  approach discussed in Refs. \cite{Ida}-\cite{Dadhich} where the brane is described as a domain wall moving in a five-dimensional bulk, which is the  $5D$ analog of the static Schwarzschild-anti-de Sitter spacetime. The consistency between our results and those in  \cite{Ida}-\cite{Dadhich} is provided by the fact that there exists a coordinate transformation that brings the $5D$ line element of the static $Sch-AdS$ bulk used in \cite{Ida}-\cite{Dadhich} (with $\Phi \neq 1$, but $\dot{\Phi} = 0$) into the  bulk in gaussian normal coordinates used in \cite{Binetruy}-\cite{Vollick} (with $\Phi = 1$ and $\dot{\Phi} = 0$). Therefore both approaches represent the {\em same} spacetime but in different coordinates \cite{mukoyama2}. Consequently, these quantities being  constant in one approach  are constant in the other one. 

This work is organized as follows. In Section 2 we set the notation and present the field equations. In Section 3 we explicitly show how that assumption $\dot{\Phi} = 0$ allows the existence of constant $\sigma$. In Section 4 we show specific models where $\sigma$ has to be a variable function of time. In Section 5 we present a summary.  

\section{Field Equations}

The equations of the brane-world paradigm can be found in a number of papers. Some authors use signature $(- + + + +)$ while other use $(+ - - - -)$. Besides, when the energy-momentum tensor is separated into  vacuum and ordinary matter,  in some works the vacuum part enters with positive sign \cite{Binetruy} while in other with negative sign \cite{Shiromizu}. Although this does not affect the final outcome, it does make difficult the comparison of intermedium formulae and the concepts of ``negative' or ``positive"  vacuum energy or tension become very entangled. 

To facilitate the discussion, and set the notation with the  ``appropriate" signs, we give a brief review of the equations. Everywhere we use signature $(+ - - - \epsilon)$. The Einstein equations in five dimensions are
\begin{equation}
\label{field equations in 5D}
{^{(5)}G}_{AB} = {^{(5)}R}_{AB} - \frac{1}{2} g_{AB}{^{(5)}R} = {k_{(5)}^2} {^{(5)}T_{AB}}, 
\end{equation}
where $k_{(5)}$ is a constant introduced for dimensional considerations and $^{(5)}T_{AB}$ is the five-dimensional energy-momentum tensor.  

These equations contain the first and second derivatives of the metric with respect to the extra coordinate. These can be expressed in terms of geometrical tensors in $4D$.

For this we introduce the normal unit ($n_{A}n^{A} = \epsilon$) vector, orthogonal to hypersurfaces $y = constant$,  

\begin{equation}
n^A = \frac{\delta^{A}_{4}}{\Phi}\;  , \;\;\;\;\;  n_{A}= (0, 0, 0, 0, \epsilon \Phi).
\end{equation}
Then, the first partial derivatives can be written in terms of the extrinsic curvature 
\begin{equation}
\label{extrinsic curvature}
K_{\alpha\beta} = \frac{1}{2}{\cal{L}}_{n}g_{\alpha\beta} = \frac{1}{2\Phi}\frac{\partial{g_{\alpha\beta}}}{\partial y},\;\;\; K_{A4} = 0,
\end{equation}
The second derivatives, $({\partial}^2g_{\mu\nu}/\partial y^2)$, can be expressed in terms of the projection $^{(5)}C_{\mu4\nu4}$ of the bulk Weyl tensor in five-dimensions, viz.,
\begin{equation}
\label{Weyl tensor}
{^{(5)}C}_{ABCD} = {^{(5)}R}_{ABCD} - \frac{2}{3}({^{(5)}R}_{A[C}g_{D]B} - {^{(5)}R}_{B[C}g_{D]A}) + \frac{1}{6}{^{(5)}R}g_{A[C}g_{D]B}.
\end{equation}
The field equations (\ref{field equations in 5D}) can be split up into three parts. In terms of the above quantities, the effective field equations in $4D$ are,
\begin{eqnarray}
\label{4D Einstein with T and K}
{^{(4)}G}_{\alpha\beta} &=& \frac{2}{3}k_{(5)}^2\left[^{(5)}T_{\alpha\beta} + (^{(5)}T^{4}_{4} - \frac{1}{4}{^{(5)}T})g_{\alpha\beta}\right] -\nonumber \\
& &\epsilon\left(K_{\alpha\lambda}K^{\lambda}_{\beta} - K_{\lambda}^{\lambda}K_{\alpha\beta}\right) + \frac{\epsilon}{2} g_{\alpha\beta}\left(K_{\lambda\rho}K^{\lambda\rho} - (K^{\lambda}_{\lambda})^2 \right) - \epsilon E_{\alpha\beta}, 
\end{eqnarray}
where
\begin{eqnarray}
E_{\alpha\beta} &=& {^{(5)}C}_{\alpha A \beta B}n^An^B\nonumber \\
&=& - \frac{1}{\Phi}\frac{\partial K_{\alpha\beta}}{\partial y} + K_{\alpha\rho}K^{\rho}_{\beta} - \epsilon \frac{\Phi_{\alpha;\beta}}{\Phi} - \epsilon \frac{k^{2}_{(5)}}{3}\left[{^{(5)}T}_{\alpha\beta} + ({^{(5)}T}^{4}_{4} - \frac{1}{2}{^{(5)}T})g_{\alpha\beta}\right].
\end{eqnarray}
Since $E_{\mu\nu}$ is traceless, the requirement $E_{\mu}^{\mu} = 0$ gives the inhomogeneous wave equation for $\Phi$, viz.,
\begin{equation}
\label{equation for Phi}
{\Phi}^{\mu}_{;\mu} = - \epsilon \frac{\partial K}{\partial y}- \Phi (\epsilon K_{\lambda \rho} K^{\lambda \rho} + {^{(5)}R}^{4}_{4}),
\end{equation}
which is equivalent to ${^{(5)}G}_{44} = k^{2}_{(5)}{^{(5)}T_{44}}$ from (\ref{field equations in 5D}). The remaining four equations are
\begin{equation}
\label{conservation equation}
D_{\mu}\left(K^{\mu}_{\alpha} - \delta^{\mu}_{\alpha}K^{\lambda}_{\lambda}\right) = k_{(5)}^2 \frac{{^{(5)}T_{4\alpha}}}{\Phi}.
\end{equation}
In the above expressions, the covariant  derivatives are calculated with respect to $g_{\alpha\beta}$, i.e., $Dg_{\alpha\beta} = 0$.

We assume that the five-dimensional energy-momentum tensor has the form
\begin{equation}
\label{AdS}
{^{(5)}T}_{AB} =  \Lambda_{(5)}g_{AB}, 
\end{equation}
where $\Lambda_{(5)}$ is the cosmological constant in the bulk. In the brane-world scenario our space-time is identified with a singular hypersurface (or $3$-brane) embedded in an $AdS_{5}$ bulk, ie., it is assumed that $\Lambda_{(5)} < 0$.

For convenience, the coordinate $y$ is chosen such that the hypersurface $\Sigma: y = 0$ coincides with the brane. Thus, the metric is continuous across $\Sigma$, but the extrinsic curvature $K_{\mu\nu}$ is discontinuous. Most brane-world models assume a ${\bf Z}_{2}$ symmetry about our brane, namely, 
\begin{eqnarray}
\label{Z2-symmetric metric}
d{\cal S}^2 &=& g_{\mu\nu}(x^{\rho}, + y)dx^{\mu}dx^{\nu} + \epsilon \Phi^2(x^{\rho}, + y) dy^2, \;\;\; for\;\; y \geq 0 \nonumber \\
d{\cal S}^2 &=& g_{\mu\nu}(x^{\rho}, - y)dx^{\mu}dx^{\nu} + \epsilon \Phi^2(x^{\rho}, - y) dy^2, \;\;\; for\;\;  y \leq 0.
\end{eqnarray}
Thus
\begin{equation}
\label{Z2 symmetry} 
K_{\mu\nu}\mid_{{\Sigma}^{+}} = - K_{\mu\nu}\mid_{{\Sigma}^{-}}.
\end{equation}
Therefore the field equations in the resulting ${\bf Z}_2$-symmetric brane universe can be written as 
\begin{equation}
\label{field equations in the Z2 universe}
{^{(5)}\bar{G}}_{AB} =  k^{2}_{(5)}(  \Lambda_{(5)} \bar{g}_{AB}+ {^{(5)}\bar{T}}_{AB}^{(brane)}),
\end{equation}
where $\bar{g}_{AB}$ is taking as in (\ref{Z2-symmetric metric}) and ${^{(5)}\bar{T}}_{AB}^{(brane)}$, with ${^{(5)}\bar{T}}_{AB}^{(brane)}n^{A} = 0$, is the energy-momentum tensor of the matter on the brane
\begin{equation}
\label{energy-momentum tensor in the brane}
{^{(5)}\bar{T}}_{AB}^{(brane)} = \delta_{A}^{\mu}\delta_{B}^{\nu}\tau_{\mu\nu}\frac{\delta(y)}{\Phi}.
\end{equation}
The delta function expresses the confinement of matter in the brane, hence
\begin{equation}
\label{brane EMT as result of integration}
\tau_{\mu\nu}(x^{\rho}, 0) = \lim_{\xi \rightarrow 0}\int_{- \xi/2}^{\xi/2}{^{(5)}\bar{T}}_{\mu\nu}^{(brane)}\Phi dy.
\end{equation}

In order to obtain the equations on the brane, we need to find an expression for the extrinsic curvature of $\Sigma$. For this, we use equation (\ref{field equations in the Z2 universe}). Since the metric is continuous across the brane, we get
\begin{equation}
\label{Evaluation of extrinsic curvature}
{^{(5)}\bar{R}}_{\mu\nu} = k_{(5)}^2\left[ - \frac{2}{3}\Lambda_{(5)} + \frac{\delta(y)}{\Phi}\left( {\tau}_{\mu\nu} - \frac{1}{3}g_{\mu\nu}\tau\right)\right].
\end{equation}
On the other hand, for the metric (\ref{metric with g44 not 1}) we have,
\begin{equation}
{^{(5)}\bar{R}}_{\mu\nu} = - \epsilon \frac{\partial}{\partial y}(\frac{K_{\mu\nu}}{\Phi}) + V_{\mu\nu},
\end{equation}
where 
\begin{equation}
V_{\mu\nu} = {^{(4)}R}_{\mu\nu} + \epsilon (2 K_{\mu\lambda}K^{\lambda}_{\nu} - K_{\mu\nu}K^{\lambda}_{\lambda}) - \frac{{^{(5)}\bigtriangledown}_{\nu} \Phi_{;\mu}}{\Phi},
\end{equation}
with $^{(5)}{\bigtriangledown}_{\nu} \Phi_{\mu} = \Phi_{\mu,\nu} - \Gamma_{\mu\nu}^{A}\Phi_{A}$.
We now substitute this into (\ref{Evaluation of extrinsic curvature}) and integrate across the brane
\begin{equation}
\lim_{\xi \rightarrow 0}\int_{- \xi/2}^{\xi/2}\left[ - \epsilon \frac{\partial}{\partial y}(\frac{K_{\mu\nu}}{\Phi})  +  V_{\mu\nu}\right]dy = k_{(5)}^2\lim_{\xi \rightarrow 0}\int_{ - \xi/2}^{\xi/2}\left[  - \frac{2 \Lambda_{(5)}}{3} +  \frac{\delta(y)}{\Phi} \left({\tau}_{\mu\nu} - \frac{1}{3}g_{\mu\nu}\tau\right)\right] dy
\end{equation}
Although the derivatives $(\partial g_{\mu\nu}/\partial y)$  and $(\partial \Phi/\partial y)$ are discontinuous across $\Sigma: y = 0$, we make the usual physical assumption that they remain finite. Thus, $\lim_{\xi \rightarrow 0} \int_{- \xi/2}^{\xi/2}V_{\mu\nu} dy = 0$, and we obtain,
\begin{equation}
\label{boundary conditions}
K_{\mu\nu}\mid_{{\Sigma}^{+}} - K_{\mu\nu}\mid_{{\Sigma}^{-}} = - \epsilon k_{(5)}^2 \left({\tau}_{\mu\nu} - \frac{1}{3}g_{\mu\nu}\tau\right),
\end{equation}
which are the usual Israel's boundary conditions. Now, from the ${\bf Z}_2$ symmetry
\begin{equation}
\label{K in terms of S}
K_{\mu\nu}\mid_{{\Sigma}^{+}} =  - K_{\mu\nu}\mid_{{\Sigma}^{-}} = - \frac{\epsilon}{2}k_{(5)}^2 \left({\tau}_{\mu\nu} - \frac{1}{3}g_{\mu\nu}\tau\right).
\end{equation}
Consequently, 
\begin{equation}
\label{emt on the brane in terms of K}
\tau_{\mu\nu} = - \frac{2\epsilon}{k_{(5)}^2}\left(K_{\mu\nu} - g_{\mu\nu} K\right).
\end{equation}
Then from (\ref{conservation equation}) and (\ref{AdS}) it follows that
\begin{equation}
\label{conservation of emt on the brane}
\tau^{\mu}_{\nu;\mu} = 0.
\end{equation}

Thus $\tau_{\mu\nu}$ represents the total, vacuum plus matter,  conserved energy-momentum tensor on the brane. It is usually separated in  two parts, 
\begin{equation}
\label{decomposition of tau}
\tau_{\mu\nu} =  \sigma g_{\mu\nu} + T_{\mu\nu},
\end{equation} 
where $\sigma$ is the tension of the brane in  $5D$, which is interpreted as the vacuum energy of the brane world, and $T_{\mu\nu}$ represents the energy-momentum tensor of ordinary matter in $4D$. 

From (\ref{K in terms of S}), (\ref{emt on the brane in terms of K}) and (\ref{decomposition of tau}) we finally get
\begin{equation}
\label{K in terms of matter in the brane}
K_{\mu\nu}\mid_{{\Sigma}^{+}} = -  \frac{\epsilon k_{(5)}^2}{2} \left(T_{\mu\nu} - \frac{1}{3}g_{\mu\nu}(T + \sigma)\right).
\end{equation}
Substituting (\ref{K in terms of matter in the brane}) and (\ref{AdS}) into (\ref{4D Einstein with T and K}), we obtain the Einstein equations with an {\em effective energy-momentum tensor} in $4D$ as
\begin{equation}
\label{EMT in brane theory}
^{(4)}G_{\mu\nu} =  {\Lambda}_{(4)}g_{\mu\nu} + 8\pi G T_{\mu\nu} - \epsilon k_{(5)}^4 \Pi_{\mu\nu} - \epsilon E_{\mu\nu},
\end{equation}
where
\begin{equation}
\label{definition of lambda}
\Lambda_{(4)} = \frac{1}{2}k_{(5)}^2\left(\Lambda_{(5)} - \epsilon \frac{ k_{(5)}^2 \sigma^2}{6}\right),
\end{equation}
\begin{equation}
\label{effective gravitational coupling}
8 \pi G =  - \epsilon \frac{k_{(5)}^4 \sigma}{6},
\end{equation}
and\footnote{With this choice of signs, for perfect fluid  $\Pi_{\mu \nu} = (1/12)[\rho^2 u_{\mu}u_{\nu}
 + \rho(\rho + 2p) h_{\mu \nu}]$ where $h_{\mu\nu} = u_{\mu}u_{\nu} - g_{\mu\nu}$.}
\begin{equation}
\label{quadratic corrections}
\Pi_{\mu\nu} =  \frac{1}{4} T_{\mu\alpha}T^{\alpha}_{\nu} - \frac{1}{12}T T_{\mu\nu} - \frac{1}{8}g_{\mu\nu}T_{\alpha\beta}T^{\alpha\beta} + \frac{1}{24}g_{\mu\nu}T^2.
\end{equation}
All these four-dimensional quantities have to be evaluated in the limit $y \rightarrow 0^{+}$.  They contain two novel features; they give a working definition of the fundamental quantities $\Lambda_{(4)}$ and $G$ and contain  higher-dimensional modifications to general relativity. Namely, local quadratic energy-momentum corrections via the tensor $\Pi_{\mu\nu}$, and the nonlocal effects from the free gravitational field in the bulk, transmitted  by $E_{\mu\nu}$.

 \subsection{Signature of the extra dimension}

The magnitude of the four-dimensional cosmological term crucially depends on the signature of the extra dimension. The  assumption $\Lambda_{(4)} = 0$, which requires $\Lambda_{(5)} = \epsilon k_{(5)}^2 \sigma^2/6$, links the sign of the extra dimension to the sign of the cosmological constant in the bulk. If we assume a spacelike (timelike) extra dimension, then the bulk must be $AdS_{5}$ $(dS_{5})$. 

The positiveness of $G$ requires $( - \epsilon \sigma) > 0$. Thus, a positive (negative) vacuum energy  $\sigma$ requires a spacelike (timelike) extra dimension. Note that $\Lambda_{(4)}$ is proportional to $\sigma^2$, so it does not depend on the actual sign of the vacuum energy. 
Finally, we note  the presence of the factor $\epsilon$ in front of the higher-dimensional corrections in (\ref{EMT in brane theory}).  Thus, the specific nature of the extra dimension, whether $\epsilon = -1$ or $\epsilon = + 1$, would play a very important role in strong gravitational fields, where these corrections dominate.

\subsection{Cosmological settings}

We now present the set of equations we need to examine the possible variation of $G$ and $\Lambda_{(4)}$. In cosmological applications the metric is commonly taken  in the form
\begin{equation}
\label{cosmological metric}
d{\cal{S}}^2 = n^2(t,y)dt^2 - a^2(t,y)\left[\frac{dr^2}{(1 - kr^2)} + r^2(d\theta^2 + \sin^2\theta d\phi^2)\right] + \epsilon \Phi^2(t, y)dy^2,
\end{equation}
where $k = 0, +1, -1$ and $t, r, \theta$ and $\phi$ are the usual coordinates for a spacetime with spherically symmetric spatial sections.

The metric coefficients are subjected to the conditions 
\begin{equation}
\label{conditions for the bulk metric on the brane}
n(t,y)|_{brane} = 1, \;\;\;a(t,y)|_{brane} = a_{0}(t).
\end{equation}
In this way the usual FLRW line element is recovered on the brane with $a_{0}$ as scale factor. The ordinary matter on the brane is usually assumed to be  a perfect fluid 
\begin{equation}
T_{\mu\nu} = (\rho + p)u_{\mu}u_{\nu} - p g_{\mu\nu},
\end{equation}
where the  energy density $\rho$ and pressure $p$ satisfy the isothermal equation of state, viz.,
\begin{equation}
\label{equation of state}
p = \gamma \rho, \;\;\;\  0 \leq \gamma \leq 1.
\end{equation}
Thus, the boundary conditions   (\ref{boundary conditions}), the ${\bf Z_{2}}$ symmetry (\ref{K in terms of S}), and (\ref{K in terms of matter in the brane})  yield
\begin{equation}
\label{density}
\rho(t) = \frac{2 \epsilon}{k_{(5)}^2 (\gamma + 1)\Phi|_{brane}}\left[\frac{a'}{a} - \frac{n'}{n}\right]_{brane},
\end{equation}
and 
\begin{equation}
\label{tension of the brane}
\sigma =   \frac{2 \epsilon}{k_{(5)}^2 (\gamma + 1)\Phi|_{brane}}\left[(3\gamma + 2)\frac{a'}{a} + \frac{n'}{n} \right]_{brane},
\end{equation}
where a prime denotes a derivative with respect to $y$. These equations show that the tension of the brane and the energy density depend on the details of the model. They  enable us to investigate the very intriguing  possibility that $\sigma$,  and consequently $G$ and $\Lambda_{(4)}$, might vary with time.  

The corresponding field equations in the bulk are
\begin{equation}
\label{G 00}
G_{0}^{0} = \frac{3}{n^2}\left(\frac{{\dot{a}}^2}{a^2} + \frac{\dot{a}\dot{\Phi}}{a \Phi}\right) + \frac{ 3 \epsilon}{\Phi^2}\left(\frac{a''}{a} + \frac{{a'}^2}{a^2} - \frac{a' \Phi'}{a \Phi}\right) + \frac{3 k}{a^2},
\end{equation}
\begin{eqnarray}
\label{G 11}
G^{1}_{1} = G^{2}_{2} = G^{3}_{3} &=& \frac{1}{n^2}\left[\frac{\ddot{\Phi}}{\Phi} + \frac{2\ddot{a}}{a} + \frac{\dot{\Phi}}{\Phi}\left(\frac{2 \dot{a}}{a} - \frac{\dot{n}}{n}\right) + \frac{\dot{a}}{a}\left(\frac{\dot{a}}{a} - \frac{2 \dot{n}}{n}\right)\right] + \nonumber \\
& & \frac{\epsilon}{\Phi^2}\left[\frac{2 a''}{a} + \frac{n''}{n} + \frac{a'}{a}\left(\frac{a'}{a} + \frac{2 n'}{n}\right) - \frac{\Phi'}{\Phi}\left(\frac{2a'}{a} + \frac{n'}{n}\right)\right] + \frac{k}{a^2},
\end{eqnarray}
\begin{equation}
\label{G zero four}
G^{0}_{4} = \frac{3}{n^2}\left(\frac{{\dot{a}}'}{a} - \frac{\dot{a} n'}{a n} - \frac{a' \dot{\Phi}}{a \Phi}\right),
\end{equation}
and 
\begin{equation}
\label{G 44}
G_{4}^{4} = \frac{3}{n^2}\left(\frac{\ddot{a}}{a} + \frac{{\dot{a}}^2}{a^2} - \frac{\dot{a}\dot{n}}{a n}\right) + \frac{ 3 \epsilon}{\Phi^2}\left(\frac{{a'}^2}{a^2} + \frac{a' n'}{a n}\right) + \frac{3 k}{a^2}, 
\end{equation}
where a prime denotes a derivative with respect to $y$.

Introducing the function \cite{Binetruy}
\begin{equation}
\label{first integral}
F(t,y) = k a^2 + \frac{(\dot{a}a)^2}{n^2} + \epsilon \frac{(a' a)^2}{\Phi^2},
\end{equation}
we find 
\begin{equation}
\label{F prime}
F' = \frac{2a' a^3}{3}k_{(5)}^2 {^{(5)}T}^{0}_{0},
\end{equation}
and
\begin{equation}
\label{F dot}
\dot{F} = \frac{2\dot{a} a^3}{3}k_{(5)}^2 {^{(5)}T}^{4}_{4},
\end{equation}
From these equations, it follows that 
\begin{equation}
\label{first integral in the bulk}
\left(\frac{\dot{a}}{na}\right)^2 =  \frac{k_{(5)}^2 \Lambda_{(5)}}{6} - \epsilon \left(\frac{a'}{a \Phi}\right)^2 - \frac{k}{a^2} + \frac{\cal{C}}{a^4},
\end{equation}
where $\cal{C}$ is a constant of integration. Evaluating this expression at the brane we obtain the generalized Friedmann equation 
\begin{equation}
\label{generalized FRLW equation}
3\left(\frac{{\dot{a}}_{0}}{a_{0}}\right)^2  =  \Lambda_{(4)} + 8\pi G \rho - \frac{\epsilon k_{(5)}^4}{12}\rho^2 - \frac{3 k}{a_{0}^2} + \frac{3 {\cal{C}}}{a_{0}^{4}}.
\end{equation}
Except for the condition that $n = 1$ at the brane, this equation is valid for arbitrary $\Phi(t,y)$ and $n(t,y)$ in the bulk \cite{Binetruy}.

\section{$\dot{\Phi} = 0$, Constant $G$ and $\Lambda_{(4)}$ }

The aim of this section is to show that all models with $\dot{\Phi} = 0$ correspond to physical situations where $G$ and $\Lambda_{(4)}$ can be taken as constants. Our discussion extends the results in Ref. \cite{Binetruy} to the case of a timelike extra dimension. In this case, without loss of generality we can set 
\begin{equation}
\Phi = 1.
\end{equation}
Then from (\ref{field equations in 5D}), (\ref{AdS}) and (\ref{G zero four})  it follows that 
\begin{equation}
\label{model with constant  Phi}
\frac{\dot{a}}{n} = \mu(t),
\end{equation}
where $\mu(t)$ is an arbitrary function. Substituting into (\ref{first integral}) and using (\ref{F prime}) we find
\begin{equation}
\label{generating equation for Phi = 1}
\epsilon (a a')' + [k + \mu^2(t)] =  \frac{k_{(5)}^2}{3}\Lambda_{(5)} a^2,
\end{equation}
 This equation can be easily integrated. 
\subsection{$\Lambda_{(5)} < 0$: $AdS_{5}$ bulk}
The signature of the extra dimension affects the evolution in time. 
For a spacelike extra dimension $(\epsilon = -1)$ we get
\begin{equation}
\label{spacelike extradimension}
a^2(t,y) = A(t)\cosh\, \omega y + B(t)\sinh\, \omega y - \frac{2[k + \mu^2(t)]}{\omega^2}.
\end{equation}

While for a timelike extra dimension $(\epsilon = 1)$,
\begin{equation}
\label{timelike extradimension}
a^2(t,y) = A(t)\cos\, \omega y + B(t)\sin\, \omega y - \frac{2[k + \mu^2(t)]}{\omega^2}.
\end{equation}
where $A(t)$ and  $B(t)$ are functions of integration and 
 \begin{equation}
\omega = \sqrt{\left( - \frac{2}{3}k_{(5)}^2 \Lambda_{(5)}\right)}.
\end{equation}
The functions $A(t)$, $B(t)$ and $\mu(t)$ are not independent. Indeed, from (\ref{conditions for the bulk metric on the brane}) we get\footnote{For simplicity we assume here that the brane is located at $y = 0$.}
\begin{eqnarray}
\mu(t) &=& {\dot{a}}_{0},\nonumber \\
\nonumber \\
A(t) &=& a_{0}^2 + \frac{2}{\omega^2}(k + {\dot{a}}_{0}^{2}).
\end{eqnarray}
From (\ref{tension of the brane}) we get the differential equation 
\begin{equation}
\label{diff. eq. for B}
a_{0}\frac{dB}{d a_{0}} + (3\gamma + 1) B = \frac{(\gamma + 1) k_{(5)}^2}{\omega}\epsilon \sigma a_{0}^{2},
\end{equation}
which can be integrated, with {\em no} contradictions if we assume constant $\sigma$, viz.,
\begin{equation}
B(t) = \frac{k_{(5)}^2 \epsilon \sigma}{3 \omega}a_{0}^{2} + b a_{0}^{- (3\gamma + 1)}, \;\;\;b = const.
\end{equation}
We note that for a  de Sitter bulk $(dS_{5})$ the situation is reversed. Namely, for a spacelike extra dimension the time-evolution of the brane in a $dS_{5}$ bulk is given by (\ref{timelike extradimension}), while for a timelike  by (\ref{spacelike extradimension}).     

\subsection{$\Lambda_{(5)} = 0$: $STM$}

This is the scenario in Space-Time-Matter theory $(STM)$ where the effective matter in $4D$ is obtained from the solutions of the $5D$ field equations with ${^{(5)}T_{AB}} = 0$. In this scenario the integration of (\ref{generating equation for Phi = 1}) yields 
\begin{equation}
\label{Liu Wesson}
a^2(t,y) = - \epsilon [k + \mu^2(t)]y^2 + C(t)y + D(t).
\end{equation}
In the context of $STM$ the functions of integration $\mu(t)$, $C(t)$, $D(t)$ are totally arbitrary. However, we can use the brane-world paradigm to evaluate them \cite{equ. STM-Brane}. Following the same procedure as above we obtain 
\begin{eqnarray}
\mu(t) &=& {\dot{a}}_{0},\nonumber \\
D(t) &=& a_{0}^2,\nonumber \\
C(t) &=& \frac{k_{(5)}^2}{3}\epsilon \sigma a_{0}^2 + c a^{- (3\gamma + 1)},\;\;\;c = const.
\end{eqnarray} 

The conclusion from the above discussion is that when $\Phi = 1$, we can {\em always} assume, with no contradictions, that the tension of the brane $\sigma$, and consequently $G$ and $\Lambda_{(5)}$, are all constants. Therefore, if the extra dimension is spacelike $(\epsilon = -1)$ by appropriately specifying the values of $\sigma$ and $\Lambda_{(5)}$ we can always set $\Lambda_{(4)} = 0$. However, this is {\em not possible} for a timelike $(\epsilon = 1)$ extra dimension\footnote{We have already mentioned that there is a coordinate transformation that brings the line element (\ref{cosmological metric}) with $\dot{\Phi} =0$ to the static $Sch-AdS$ form. Therefore this also holds for the approach where the brane is described as a domain wall moving in a $Sch-AdS$ bulk used in Refs. \cite{Ida}-\cite{Dadhich}.}.

For a constant $\sigma$, it follows from  (\ref{conservation of emt on the brane}) and (\ref{decomposition of tau})  that $T^{\mu}_{\nu;\mu} = 0$. For perfect fluid satisfying the equation of state (\ref{equation of state}) this leads to $\rho = \rho_{c}a_{0}^{- 3(\gamma + 1)}$, with $\rho_{c} = const$. Substituting this into (\ref{generalized FRLW equation}) we obtain the equation that governs the evolution of the scale factor $a_{0}$ and  defines the rest  of the functions of time appearing in the problem.

We would like to emphasize that in the above solutions the extra dimension may be either spacelike or timelike. Therefore, they extend the results of \cite{Binetruy}. We should also  mention that  the $STM$ solution given by (\ref{model with constant  Phi}) and (\ref{Liu Wesson}) with $\epsilon = -1$   has recently been discussed in the literature, although in another context \cite{LiuWesson}.

\section{$\dot{\Phi} \neq 0$, Variable $G$ and $\Lambda_{(4)}$ }

Here we show some models for which the introduction of a non-static $\Phi$ leads to physical scenarios where $G$ and $\Lambda_{(4)}$ {\em cannot} be taken as constants but are varying functions of time.

\subsection{Models with $\bf{n' = 0}$}

The first model is generated by the assumption that  $n' = 0$. That is we assume that we can set 
\begin{equation}n = 1
\end{equation}
everywhere, not only on the brane. 
Then, from $G_{04} = 0$ we get
\begin{equation}
\label{case n constant}
\frac{a'}{\Phi}= h(y)
\end{equation}
where $h(y)$ is an arbitrary function. Substituting this into (\ref{first integral}) and using (\ref{F dot})  we find
\begin{equation}
\label{generating equation for n const case}
\frac{d({\dot{a}a})}{dt} - \frac{k_{(5)}^2}{3}\Lambda_{(5)}a^2 + [k + \epsilon h^2(y)] = 0.
\end{equation}
Integrating we obtain:
\subsubsection{$AdS_{5}$}
In the present model the time-evolution of the universe is {\em not} affected by the signature of the extra dimension, contrary to what happens in the previous $(\Phi = 1)$ case. Thus, 
\begin{equation}
\label{AdS with n constant}
a^2(t,y) = f(y)\sin\, \omega t + g(y)\cos\, \omega t - \frac{2[k + \epsilon h^2(y)]}{\omega^2},
\end{equation}
where $f(y)$ and $g(y)$ are arbitrary functions of integration. 
Substituting the above into (\ref{first integral in the bulk}) we find a relationship between the functions, namely
\begin{equation}
f^2(y) + g^2(y) = \frac{4}{\omega^2}\left( {\cal{C}} + \frac{[k + \epsilon h^2(y)]}{\omega^2}\right).
\end{equation}
The fulfillment of this condition assures that the above is a solution of all Einstein's equations (\ref{G 00})-(\ref{G 44}).

We note that in the case of a de Sitter bulk $(\Lambda_{(5)} > 0)$ we recover the familiar exponential expansion in time.

\subsubsection{$STM$} 
Setting $\Lambda_{(5)} = 0$ in (\ref{generating equation for n const case}), we obtain
\begin{equation}
\label{n constant with zero lambda}
a^2(t,y) = - [k + \epsilon h^2(y)]t^2 + l(y)t + m(y),
\end{equation}
where $l(y)$ and $m(y)$ are functions of integration. From (\ref{first integral in the bulk}) we find that they should satisfy the relation 
\begin{equation}
\frac{l^2(y)}{4} + [k + \epsilon h^2(y)] = \cal{C}. 
\end{equation}
When this condition is met,  the generalized Friedmann equation (\ref{generalized FRLW equation}) and (\ref{G 00})-(\ref{G 44}) are identically satisfied. We note that some cosmological implications of this solution are discussed in Ref. \cite{Fukui Seahra and Wesson}. 

\subsubsection{Variable $\sigma$} 
From (\ref{tension of the brane}) we get
\begin{equation}
\sigma = \frac{2(3\gamma + 2)}{k_{(5)}^2 (\gamma +1)} \frac{\epsilon h(0)}{a_{0}(t)},
\end{equation}
where $a_{0}(t)$ is one of the functions (\ref{AdS with n constant}) or (\ref{n constant with zero lambda}) evaluated at the brane.  Thus, in the present model the tension cannot be a  constant, instead it is  {\em always} a varying function. Then from (\ref{definition of lambda}) it follows that $\Lambda_{(4)} \sim a_{0}^{-2}$ and from (\ref{effective gravitational coupling})
\begin{equation}
8\pi G =  - \frac{k_{(5)}^2(3\gamma + 2)}{3(\gamma + 1)} \frac{h(0)}{a_{0}(t)}.
\end{equation}
Positiveness of $G$ requires $h(0) < 0$. On the other hand, from (\ref{density}) we find that $\rho > 0$ demands $\epsilon h(0) > 0$. This is the only requirement on $h(y)$, otherwise it is arbitrary. For example, the choice $h(y) = - e^{- |y|}$ meets the requirements for $\epsilon = -1$. Therefore, the present model with varying $\Lambda_{(4)}$ and $G$ works well for a spacelike extra dimension. 

Now we can formulate some general conclusions about the behavior of the fundamental quantities $G$ and $\Lambda_{(4)}$. Namely,  the question of whether they change in time or not, is independent of (i) the specific value and sign of the cosmological constant in $5D$, and (ii) the signature of the extra dimension.   

\subsection{Separable models}

We have seen that each of the conditions $\dot{\Phi} = 0$ and $n' = 0$ independently  allows the complete integration of the field equations. The similar condition $a' = 0$ $(\dot{a} \neq 0)$ requires  $n' = 0$.  Thus $a' = 0$ leads to an empty brane; $\sigma = \rho = 0$. 

Here we will consider the class of models for which the metric coefficients are separable functions of their arguments. Without loss of generality we can set
\begin{equation}
n = n(y), \;\;\; a(t,y) = a_{0}(t)Y(y), \;\;\;\; \Phi = \Phi(t).
\end{equation}
From $G_{04} = 0$ it follows that 
\begin{equation}
\left(\frac{n'}{n}\right) = \beta \left(\frac{Y'}{Y}\right),\;\;\;\frac{\dot{\Phi}}{\Phi} = (1 - \beta)\frac{\dot{a}}{a},
\end{equation}
where $\beta$ is a separation constant. Thus from (\ref{tension of the brane}) we find
\begin{equation}
\sigma = \frac{2(3\gamma + \beta + 2)}{k_{(5)}^2 (\gamma + 1)} \left(\frac{Y'_{brane}}{Y_{brane}}\right)\frac{\epsilon}{\Phi(t)}. 
\end{equation}
Thus, the general feature of {\em all} these models is that $\sigma$ is a function of time, unless the fifth dimension is static, i.e., $\Phi$ is constant. This corresponds either to a motionless universe or to an empty space, which we disregard in the present study. 

\subsubsection{The standard cosmological model}

Here in order to illustrate this general feature, we consider the five-dimensional metric \cite{JPdeL 1} 
\begin{equation}
\label{Ponce de Leon solution}
d{\cal S}^2 = y^2 dt^2 - t^{2/\alpha}y^{2/(1 - \alpha)}[dr^2 + r^2(d\theta^2 + \sin^2\theta d\phi^2)] - \alpha^2(1- \alpha)^{-2} t^2 dy^2,
\end{equation}
where $\alpha$ is a constant. This is a solution to the five-dimensional field equations (\ref{field equations in 5D}), (\ref{G 00})-(\ref{G 44}) with ${^{(5)}T}_{AB} = 0$. Therefore, it is usually discussed in the context of $STM$.

In four-dimensions (on the hypersurfaces $y = const.$) this metric corresponds to the $4D$ Friedmann-Robertson-Walker models with flat $3D$ sections. The  energy density $\rho_{eff}$  and pressure $p_{eff}$ of the effective $4D$ matter, defined by the right hand side of (\ref{EMT in brane theory}), satisfy the equation of state
\begin{equation} 
\label{eq of state for the eff fluid}
p_{eff} = n \rho_{eff},  
\end{equation}
where $n = ({2\alpha}/{3} -1)$. Thus  for $\alpha = 2$ we recover radiation, for $\alpha = 3/2$  we recover dust, etc. This solution, which is usually called {\em standard} $5D$ cosmological model in $STM$ \cite{Wesson book}, has been applied to the discussion of a wide variety of cosmological problems that range from singularities to geodesic motion \cite{Fukui Seahra and Wesson}, \cite{Overduin}-\cite{Seahra 2}.

Here we will use the brane-world paradigm to take a closer look on the effective matter quantities $\rho_{eff}$ and $p_{eff}$. They are the same in both, $STM$ and brane theories. However, in $STM$ there is no enough information as to know, and evaluate, the different contributions that make up these effective quantities. 

In a recent paper we discussed the ``equivalence" between  brane and $STM$ theories \cite{equ. STM-Brane}. Here, we will use (\ref{Ponce de Leon solution}) as the generating $5D$ space for brane-world models. First we have to construct the ${\bf{Z}}_{2}$ symmetric brane. 

Now we cannot set the brane at $y = 0$. We  set it at $y = y_{0}$ and impose the ${\bf Z}_{2}$ symmetry under the transformation $y \rightarrow y_{0}^{2}/y$ \cite{Grojean}. The appropriate bulk background is 
\begin{equation}
\label{Bulk +}
d{\cal S}^2 = \frac{y^2}{y_{0}^{2}} dt^2 - t^{2/\alpha}\left(\frac{y}{y_{0}}\right)^{2/(1 - \alpha)}[dr^2 + r^2(d\theta^2 + \sin^2\theta d\phi^2)] - \frac{\alpha^2}{y_{0}^2(1- \alpha)^2}
t^2 dy^2,
\end{equation}
\begin{equation}
\label{Bulk -}
d{\cal S}^2 = \frac{y_{0}}{y^2} dt^2 - t^{2/\alpha}\left(\frac{y_{0}}{y}\right)^{2/(1 - \alpha)}[dr^2 + r^2(d\theta^2 + \sin^2\theta d\phi^2)] - \frac{\alpha^2}{(1- \alpha)^2}\frac{y_{0}^{2}}{y^4} t^2 dy^2,
\end{equation}
for $y \geq y_{0}$ and $y \leq y_{0}$, respectively. Here the extra dimension is spacelike ($\epsilon = -1$). The induced $4D$ metric on the brane, at $y = y_{0}$, is the Robertson-Walker metric,
\begin{equation}
\label{Robertson Walker}
ds^2 = dt^2 - t^{2/\alpha}[dr^2 + r^2(d\theta^2 + \sin^2\theta d\phi^2)].
\end{equation}
From (\ref{extrinsic curvature}) and (\ref{Bulk +}) we find
\begin{eqnarray}
\label{K +}
K^{t}_{t} &=& \frac{|1 - \alpha|}{|\alpha|} \frac{y_{0}}{y t},\nonumber \\
K^{r}_{r} &=& K^{\theta}_{\theta} = K^{\phi}_{\phi} = \frac{|1 - \alpha|}{|\alpha|(1 - \alpha)}\frac{y_{0}}{ y t}.
\end{eqnarray}
Using (\ref{Bulk -}) we obtain the same expressions as in (\ref{K +}) but with opposite sign, as one expected. We now use (\ref{K in terms of matter in the brane}) and obtain (here $\epsilon = -1$)
\begin{eqnarray}
2\rho + 3p  - \sigma &=& \frac{6|1 - \alpha|}{|\alpha| k_{(5)}^{2}}\frac{1}{t},\nonumber \\
 \rho + \sigma &=& \frac{6|1 - \alpha|}{|\alpha|( \alpha - 1) k_{(5)}^{2}}\frac{1}{t}.
\end{eqnarray}
The physical condition $\tau_{0}^{0} = (\sigma + \rho) > 0$ puts a lower limit on $\alpha$, namely  $\alpha > 1$. It is easy to verify that this requirement also assures the fulfillment of $(\rho + p) > 0$.  Now  using the equation of state (\ref{equation of state}) we get
\begin{equation}
\label{matter density}
\rho = \frac{2}{k_{(5)}^2 (\gamma + 1)t},
\end{equation}
and
\begin{equation}
\label{sigma}
\sigma =  \frac{2[3(\gamma + 1) - \alpha]}{k_{(5)}^{2} \alpha (\gamma + 1) t}.
\end{equation}
\subsubsection{Variable $G$ and $\Lambda_{(4)}$}
Consequently, the effective gravitational ``constant" $G$ varies with time as $G \sim t^{-1}$. Namely, 
\begin{equation}
\label{Var of G from standard cosmology}
8\pi G = \frac{k_{(5)}^2[3(\gamma + 1) - \alpha]}{3 \alpha (\gamma + 1) t}.
\end{equation}
The positiveness of $G$ puts an upper limit on $\alpha$, namely $\alpha < 3(\gamma + 1)$. Thus 
\begin{equation}
1 < \alpha < 3(1 + \gamma),
\end{equation}
which for the effective quantities implies $- 1/3 < n < 3(1 + \gamma)$. Consequently, for all ``allowed" values $(\rho_{eff} + 3p_{eff}) > 0$. We note however that in order to explain the modern observations that our universe is expanding  with positive acceleration this energy condition should be violated\footnote{For $\alpha < 1$ the models describe inflationary situations where the cosmological fluid has repulsive properties \cite{Seahra 1}, \cite{Seahra 2}. The time variation of $G$ is usually written as $(\dot{G}/G) = g H$, where $g$ is a dimensionless parameter, whose present observational upper bound is $|g| \leq 0.1$ \cite{Melnikov1}-\cite{Melnikov3}. Here from (\ref{Var of G  from standard cosmology}) we obtain $(\dot{G}/G) = - \alpha H$, it follows that the dominant and strong energy conditions should be violated.}.

We also obtain a time-epoch-varying effective cosmological term, namely 
\begin{equation} 
\Lambda_{(4)} =  \frac{[3(\gamma + 1) - \alpha]^2}{\alpha^2 (\gamma + 1 )^2 t^2} =  \frac{[3\gamma + 2 - q]^2}{(\gamma + 1 )^2 } H^2,
\end{equation}
where $H = (\dot{a}/a) = (\alpha t)^{- 1}$ and $q = - (a \ddot{a}/{\dot{a}}^2) = \alpha -1$ are  the Hubble and ``deceleration" $4D $ parameters, respectively. We observe that $\Lambda_{(4)}$ makes a positive contribution to the effective energy density defined by (\ref{EMT in brane theory}). Notice that we cannot set $\Lambda_{(4)} = 0$; this would lead to the unphysical requirement of $G = 0$.

We would like to finish this section with the following comments:

 (i) We note that $G\rho \sim t^{- 2}$. Thus, although $\rho$ and $G$ do depend on the unknown five-dimensional constant $k_{(5)}$, the observational consequences in cosmology do not. 

(ii) There is an extensive literature suggesting that the relation $\Lambda_{(4)} \sim H^2$ plays a fundamental role in cosmology. More recently, this relation was obtained from a model based on the quantum gravitational uncertainty principle and the discrete structure of spacetime at Plank length  \cite{Vishwakarma}. The dependence $\Lambda_{(4)} \sim H^2$ explains the current observations successfully and provides a much needed large age of the universe. 

(iii) By virtue of the variation of $\sigma$ the energy-momentum tensor $T_{\mu\nu}$, describing the ordinary matter on the brane, is not conserved separately. What is conserved here are the effective energy density and pressure obtained after adding all the contributions in  (\ref{EMT in brane theory}). Only these effective (or total) quantities have observational consequences.

(iv) The  model discussed above is intended to be illustrative of what we can expect from the theory rather than an experimental/observational suggestion. The study of other, more realistic, models is necessary.

\section{Conclusions}

The physical scenario of variable $\Phi$ cannot be ruled out {\em a priori}. The constancy of this function is an external condition and not a requirement of the field equations.

When ${\dot{\Phi} = 0}$, the cosmological equations discussed in Sec. 2.2 have solutions on the brane (fixed at some $y = constant$) that are consistent with the assumption of constant $\sigma$.  The same is true in the approach where the brane is modeled as a domain wall moving in a $Sch-AdS$ bulk. As mentioned earlier, there is a coordinate transformation that brings the bulk metric (\ref{cosmological metric}) with static ${\Phi}$ into the static $Sch-AdS$ form. 

Thus, our solutions with $\dot{\Phi} \neq 0$ do not contradict previous results which assume a  static fifth dimension. We have given two specific models where the physical entities $\sigma$, $G$ and $\Lambda_{(4)}$ are not just parameters in the theory, but are dynamical quantities that evolve with time. The first model arises  from the assumption of $n' = 0$, while the other requires  separation of variables. Each of these assumptions leads to a complete integration of the field equations in $5D$ which correspond to distributions of matter in $4D$ that satisfy usual physical requirements as the energy conditions and positiveness of $G$. 

These models are very simple, but they clearly illustrate our main point here. Namely, that the introduction of a varying $\Phi$ in brane models  entails the  possibility  of variable fundamental physical ``constants". The study of other models of this kind is important\footnote{Other models of this kind are the solutions to the cosmological equations in Section 2.2 obtained under the assumption of ``self-similar" symmetry. We leave the discussion of such models to a future work.}. This would give us the opportunity to test them for compatibility with observational data.

We should emphasize that the behavior of $\sigma$, $G$ and $\Lambda_{(4)}$ is independent of the specific value or sign of the cosmological constant in $5D$. In addition, the specific nature on the extra dimension may (as in the $\dot{\Phi} = 0$ case ) or may not (as in the $n' =0$, $\dot{\Phi} \neq 0$ case ) influence the time evolution of the brane.  

It is well known that variations of fundamental physical ``constants", in particular of $G$, are predicted in numerous scalar-tensor theories, and multidimensional theories. However, our approach and results are {\em not} equivalent to these theories. Indeed, in Jordan-Brans-Dicke type theories $G$ does depend on the scalar field, and in multidimensional theories $G \sim \Phi^{- N}$, where $N$ is the number of extra dimensions \cite{Melnikov1}-\cite{Melnikov3}. However, these predictions make no reference to the cosmological term $\Lambda_{(4)}$ and/or $\sigma$; the variation of $G$ does not seem to affect them. This is notoriously different in the context of brane-world models where the variations of these quantities are related.

\end{document}